\documentclass{INTERSPEECH2023}
\usepackage{multicol,multirow}
\usepackage{url,cite}
\usepackage{hyperref}
\usepackage{xcolor}
\usepackage{tabularx,lipsum}
\usepackage{ctable}
\usepackage{float}
\usepackage{balance,subcaption}
\usepackage{comment}
\usepackage{svg}
\usepackage{censor}
\hypersetup{
    colorlinks=true,
    linkcolor=purple,
    filecolor=magenta,      
    urlcolor=purple,
}
\interspeechcameraready 

\newcommand{\newpara}[1]{\vspace{8pt}\noindent\textbf{#1}}
\newcolumntype{Y}{>{\centering\arraybackslash}X}

\title{Malafide: a novel adversarial convolutive noise attack against \\ deepfake and spoofing detection systems}
\name{Michele Panariello$^{*}$, Wanying Ge$^{*}$\thanks{$^{*}$ These authors contributed equally to this work.}, Hemlata Tak, Massimiliano Todisco and Nicholas Evans}

\address{EURECOM, Sophia Antipolis, France}
\email{firstname.lastname@eurecom.fr}

\newcommand{\spf}{\mathbf{s}}
\newcommand{\spfset}{\MakeUppercase{{\spf}}}

\DeclareMathOperator{\countermeasure}{CM}
\newcommand{\cm}[1]{\countermeasure(#1)}
\newcommand{\advflt}{\mathbf{m}}
\newcommand{\utt}{\mathbf{u}}
\DeclareMathOperator{\uniform}{U}

\begin{document}

\maketitle
 
\begin{abstract}
We present \emph{Malafide}, a universal adversarial attack against automatic speaker verification (ASV) spoofing countermeasures (CMs).
By introducing convolutional noise using an optimised linear time-invariant filter, Malafide attacks % sttsvld
can be used to compromise CM reliability
while preserving other speech attributes such as quality and the speaker's voice. 
In contrast to other adversarial attacks proposed recently, 
Malafide filters are optimised independently of the input utterance and duration, are tuned instead to the underlying spoofing attack, and require the optimisation of only a small number of filter coefficients.
Even so, they degrade CM performance estimates by an order of magnitude, even in black-box settings, and can also be configured to overcome integrated CM and ASV subsystems.  
Integrated solutions that use self-supervised learning CMs, however, are more robust, under both black-box and white-box settings.
\end{abstract}
\vspace{-0.1cm}
\section{Introduction}
Spoofing, or presentation attacks can be used by a fraudster to
% adversarial
manipulate the behaviour of a biometric recognition system and hence to gain illegitimate access to protected systems, services or facilities.
Auxiliary sub-systems in the form of countermeasures (CMs)
are nowadays commonly deployed in order to defend against such attacks and can offer strong levels of protection~\cite{tolosana2019biometricPAD,george2019FacePAD}, including in the case of automatic speaker verification (ASV)~\cite{yamagishi2021_ASV_spoof,jungsasv2022}, the focus in this paper. 
CMs aim to detect tell-tale signs of spoofing, namely processing artefacts that are not expected in recordings of bona fide, genuine human speech.

Spoofing and detection research are nonetheless a game of \emph{cat and mouse} 
in which a defender continually adapts to emerging threats, while an attacker or adversary continually adapts to the resulting CMs.
While threats to ASV have evolved in the form of ever-more-effective synthetic speech and converted voice attacks, CMs have largely kept apace~\cite{wang22_ssl,tak2022automatic, jung2022aasist}.
New threats nonetheless continue to emerge.

With defences now commonplace, adversaries can adapt to conceal the artefacts which might otherwise serve to distinguish bona fide from spoofed biometric samples;
they can devise new attacks to manipulate not just the biometric classifier, but also a spoofing CM.
Examples of such adversarial attacks have already emerged~\cite{cm_adv_atks1,black_box_cm_additive,gomezalanis2021adversarial,universal_perturbations_CM,adv_kassis,wu2020defense}.
Most take the form of additive noise whereby utterance-specific perturbations are learned and \emph{added} to a speech signal.
These approaches cannot be implemented in real time and are sensitive to the specific utterance;
the perturbations tend to be highly sensitive and even slight distortions can render the attack ineffective~\cite{neekhara19b_universal,xie2020real}. 

In this paper, we propose \emph{Malafide}, a novel attack designed for the strength-testing of spoofing and deepfake detection solutions against adversarial, \emph{convolutive} noise attacks. 
Malafide attacks involve the optimisation of a linear time-invariant filter which is applied to deepfake or spoofed utterances in order to provoke their misclassification as bona fide utterances. 
Convolutive noise is independent of an utterance and its duration and is naturally robust to time-domain shifts, unlike additive adversarial noise. 
Malafide attacks require the optimisation of only a small number of filter coefficients, far less than the number of waveform samples or, equivalently, 
the number of samples that would need to be generated in an additive noise attack.
They are also more \emph{universal} than additive noise attacks in that they are not optimised for each utterance or speaker, but are instead optimised for a given spoofing attack, with the latter acting to ensure they compromise both CM and ASV subsystems. 
The attack is optimised offline and can hence be applied in real time.

While attack studies of the nature discussed above can raise obvious ethical questions, such work is key to the typical adversarial development cycle.  
Only by continually probing and strength-testing a given system and by addressing any identified weaknesses can there be any confidence in its security.
This is the spirit of our work presented in this paper.

\section{Relation to prior work}
Adversarial attacks, first introduced for image related tasks~\cite{adv_attacks_goodfellow}, % were 
have also been studied in the speech domain, e.g.\ for automatic speech recognition (ASR)~\cite{audio_adv_attacks_carlini,neekhara19b_universal} as well as %being applied in studies of
spoofing and ASV~\cite{cm_adv_atks1, xie2020real,black_box_cm_additive,universal_perturbations_CM,wu2020defense,adv_kassis}.
Early studies~\cite{cm_adv_atks1, adv_attacks_goodfellow} explored 
adversarial examples in 
the form 
of additive noise 
and 
attack transferability 
in black-box scenarios.
Typically, these attacks operate on feature representations, implying that the attacker has corresponding system-level access.  
This is unlikely in practice and the attack threat diminishes greatly if time-domain speech signals are resynthesized from features~\cite{adv_kassis}.
Malafide attacks 
are applied at the level of raw speech signals in the time domain.

The early approaches consider the generation of utterance-specific adversarial noise.
Again drawing upon inspiration from studies in image processing~\cite{universal_perturbations_image}, universal adversarial perturbations
have been adapted to a wide variety of audio tasks, such as ASR~\cite{universal_perturbations_ASR}, ASV~\cite{unviersal_perturbations_ASV}, and environmental sound classification~\cite{universal_audio_attacks_ASC}.
Common to these approaches is the generation of adversarial perturbations by 
iteratively optimising over several data points.  
We adopt a similar approach, but generate %to that 
adversarial noise that is specific to an underlying spoofing attack.

To the best of our knowledge, the most relevant prior work is~\cite{universal_perturbations_CM}, which reports an investigation of universal perturbations against spoofing and deepfake CM systems.  This technique, though, targets the simultaneous manipulation of both CM and ASV subsystems in a manner that is independent to specific spoofing attacks.
So that they compromise the ASV system, perturbations are also generated for specific speakers, which adds complexity and precludes their usage on unseen speakers. Moreover, there is no explicit constraint that acts to protect speech quality.
Our approach differs in that it operates alongside specific spoofing attacks to augment the threat they pose to combined CM and ASV subsystems.  
We assume that the spoofing attack is sufficient to manipulate the ASV subsystem but, by protecting other speech attributes such as intelligibility, prosody and the speaker's voice, 
Malafide attacks act to compromise both CM and ASV subsystems.
Different to all previous work, our approach involves the learning of an aversarial linear, time-invariant (LTI) filter which can be applied in real-time to a spoofed utterance through time domain convolution.

\section{Malafide attacks}
\label{sec:attack_model}
Let $\spfset^{(a)} = \{\spf_1^{(a)}, \spf_2^{(a)} \dots \spf_N^{(a)}\}$ be a set of deepfake/spoofed utterances generated by algorithm $a$ (a particular text-to-speech or voice conversion algorithm). 
Spoofed utterances are generated to manipulate an ASV system so as to increase the likelihood of it verifying erroneously claimed identities. 
Auxiliary detection classifiers in the form of CMs are used to defend against spoofing attacks and hence to protect ASV reliability.
Let $\cm{\utt} = s\left(y\mid\utt\right)$ %\textcolor{red}{
be a model that assigns a score $y$ to utterance $\utt$ where, by convention, higher scores reflect greater support for the bona fide class and lower scores greater support for the spoof class.
Ideally, for most spoofed utterances~$i$, $\cm{\spf_i^{(a)}}$ will produce low scores.

\begin{figure}[!t]
    \centering
    \includegraphics[width=\columnwidth]{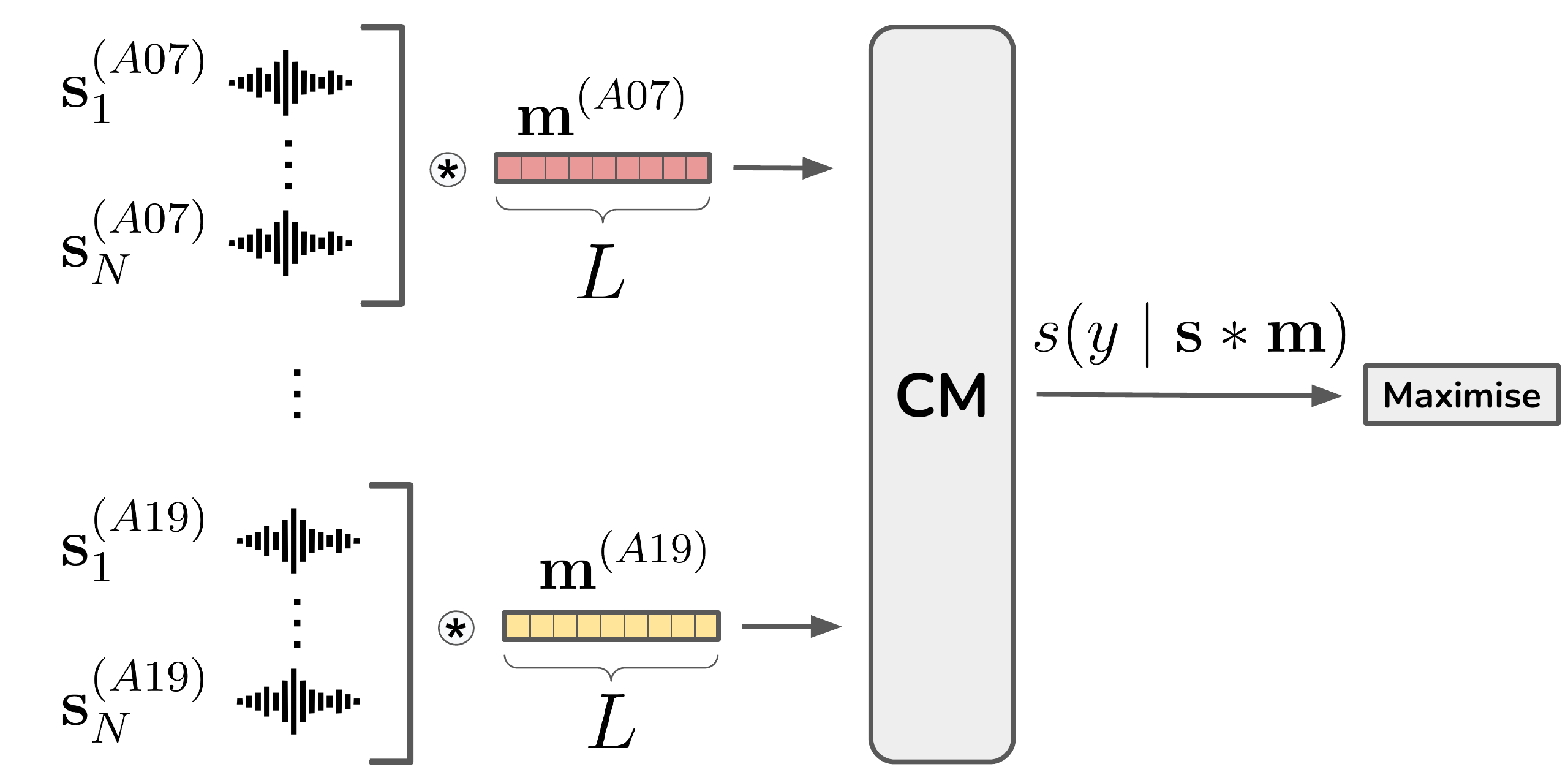}
    \caption{The Malafide filter training procedure.}
    \label{fig:training}
\end{figure}

Malafide attacks involve the optimisation of a linear time-invariant (LTI), non-causal filter, the coefficients (tap weights) of which are optimised to provoke the misclassification of deepfake/spoofed utterances as bona fide utterances.
The LTI, $L$-tap filter $\advflt^{(a)}$ is designed to maximise $\cm{\spf_i^{(a)} * \advflt^{(a)}}$ (where $*$~denotes convolution).
In the case of several different spoofing algorithms $a_1 \dots a_K$, an attacker can optimise an equivalent number of filters $\advflt^{(a_1)} \dots \advflt^{(a_K)}$. 
The learning of attack-specific filters is intuitive given that spoofing artefacts are also attack-specific.
The filter should then be tuned to counter the reliance of the CM upon these same artefacts.

Filter coefficients $\advflt^{(a)}$ can be optimised with conventional gradient descent using the set of spoofed utterances $\spfset^{(a)}$. The objective function is given by
\begin{equation}
    \label{eq:objective}
    \max_{\advflt^{(a)}} \sum_i \cm{\spf_i^{(a)} * \advflt^{(a)}}
\end{equation}
A graphical depiction of the training procedure is shown in Figure~\ref{fig:training} for two arbitrary spoofing attacks.  An $L$-tap filter is optimised separately for each attack to manipulate the behaviour of a common CM.

Without constraints, Malafide filtering can cause excessive speech degradation. %result in degraded speech quality.
For detection settings in the absence of an ASV system or a human listener (e.g.\ a standalone CM operating to detect deepfakes), this may have little consequence.  
Where the CM is deployed alongside an ASV system, however, the filter may act to compromise the CM, but might introduce distortion of sufficient level that 
the spoofing attack 
is no longer successful in compromising the ASV system. 
Accordingly, 
$\advflt^{(a)}$ should be constrained somehow so as to strike a balance between the maximisation of~\eqref{eq:objective} and the preservation speech fidelity,
e.g.\ intelligibility, prosody or the speaker's voice.

We have found that such a suitable balance can be achieved by initialising
$\advflt$ to \emph{resemble} a convolutive identity, i.e.\ an impulse response which exhibits a dominant Dirac (delta) function.
We use
He initialization~\cite{he_init} whereby
each filter coefficient
is set to some random value sampled from a uniform distribution $r\sim\uniform(-\sqrt{3 / L},\sqrt{3 / L})$.
The central coefficient of $\advflt$ at $t=0$ is then set to $1$.
The filter coefficients are optimised via gradient descent according to \eqref{eq:objective} but, to preserve the Dirac property, the central coefficient is reset to $1$ after
each filter update derived from a batch.
The number of taps, or filter length $L$ provides an additional level of control over the balance between the preservation of speech fidelity and the effectiveness of the attack.
Filters with a longer impulse response allow for greater control or manipulation and hence stronger attacks, but introduce greater distortion. 
Shorter impulse responses produce less distortion, but also weaker attacks.

\begin{figure}[!t]
    \centering
    \includegraphics[width=1\columnwidth]{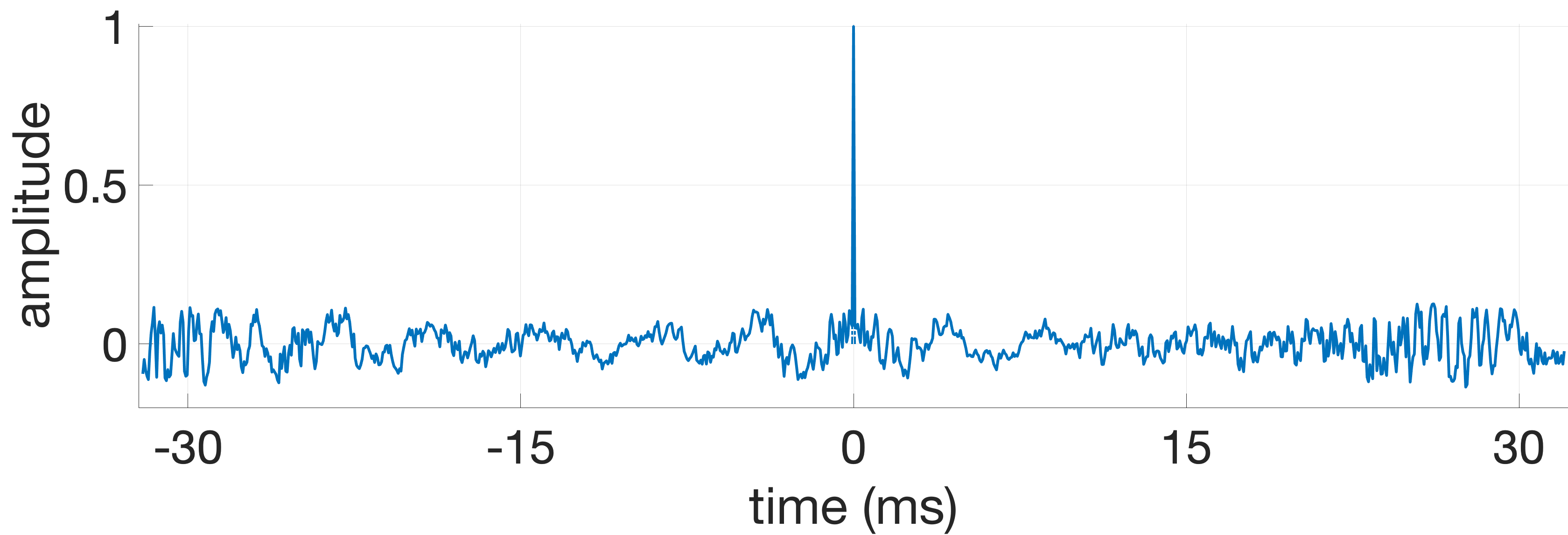}
    \includegraphics[width=1\columnwidth]{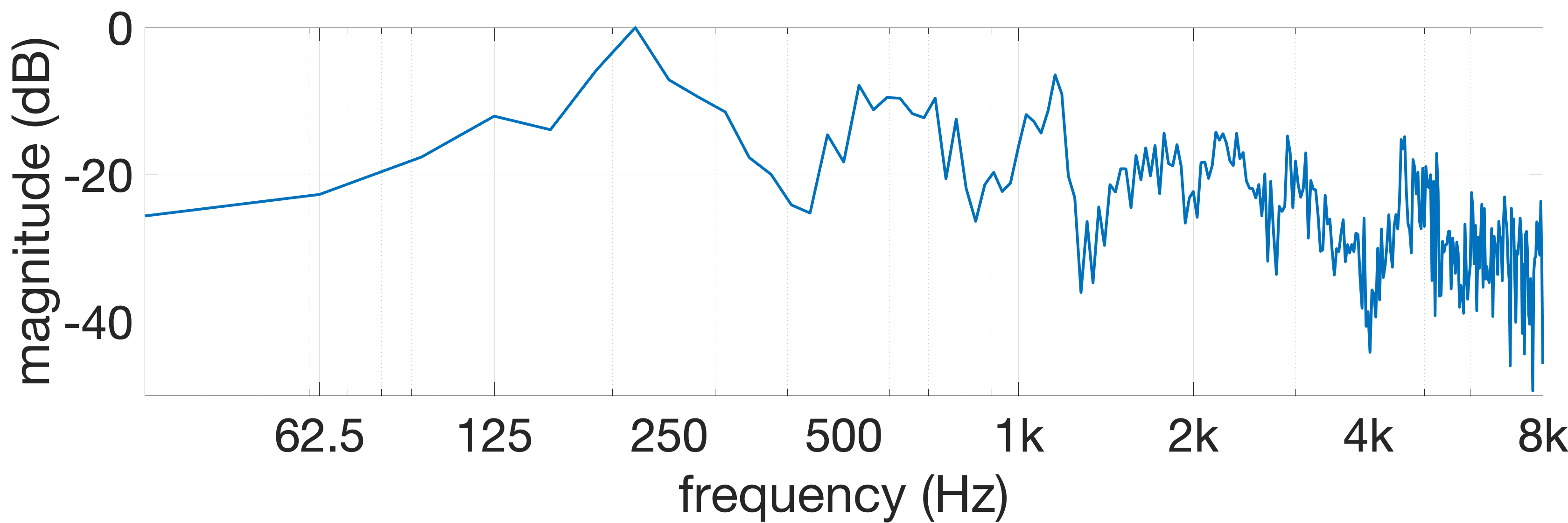}
    \caption{ Impulse (top) and normalised magnitude frequency (bottom) responses
    for a Malafide filter optimised for a 
    RawNet2 CM and A10 spoofing attacks.}
    \label{fig:mala}
\end{figure}

The impulse response of a 1025-tap Malafide filter optimised for an arbitrarily selected A10 spoofing attack and a RawNet2 CM is illustrated to the top of Figure~\ref{fig:mala}. The non-causal filter reflects the Dirac property at $t=0$, with lower, off-centre coefficients. 
The corresponding normalised magnitude frequency response is illustrated on a decibel magnitude, log frequency scale to the bottom of Figure~\ref{fig:mala}. 
It shows pronounced attenuation around 450~Hz, 900~Hz, 1.3~kHz, 4~kHz and 8~kHz, an indication of where the RawNet2 CM \emph{sees} A10 spoofing artefacts. 
By suppressing these frequency intervals, the filter acts to suppress the artefacts which the CM otherwise uses to distinguish bona fide from spoofed utterances.

\section{Experimental setup}

\subsection{Protocols and filter optimisation}
\label{sec:protocols}
All experiments were conducted using the ASVspoof 2019 logical access (LA) dataset~\cite{wang2020ASVspoof}. It contains spoofing attacks generated with a set of 
algorithms labelled 
A01 to A19.
Attacks A01 to A06 are contained in both the \emph{training} and \emph{development} partitions, while A07 to A19 are contained only in the \emph{evaluation} partition.
Training and development partitions relate to the realm of a defender whose role is to train and develop spoofing CMs. 
The set of three CMs described in Section~\ref{sec:cms} are trained in the usual way using these two data partitions.

In contrast, the test partition contains data in the realm of the attacker.
Attack-specific filters are hence trained according to~\eqref{eq:objective} using a subset of the test partition, i.e.\ using A7 to A19 spoofing attack data. 
We stress that, in contrast to usual practice, the use of \emph{test} data for \emph{training} purposes is acceptable in this case; the attacker is not bound by experimental protocols and can use test data in any reasonable way that is to their advantage.
The test partition is nonetheless split 
into two parts which contain an equal number of utterances for each attack.
Attack-specific adversarial filters are then optimised using only data in Part~1
and tested using data in Part~2.

This re-partitioning allows us to verify the \emph{universality} of Malafide attacks, namely the effectiveness or transferability of the attack to unseen utterances. 
The setup reflects a scenario in which filters are trained by an attack offline and then used to implement online/real-time attacks, e.g.\ in a logical access or telephony scenario.
While all data partitions contain both bona fide and spoofed data,
Malafide filter optimisation is performed using spoofed utterances only.

\subsection{Implementation}
\eqref{eq:objective} is optimized with Adam~\cite{kingma2014adam}.
The learning rate and weight decay are tuned separately for each CM model. 
Filters are optimised for $15$ epochs with a batch size of $14$ using data in Part~1 (see Section~\ref{sec:protocols}).
During optimisation of \eqref{eq:objective}, the weights of the CM model are kept frozen.
We explored different filter lengths $L$ (65, 129, 257, 513, 1025, 2049 and 4097) in order to explore the balance between optimisation of~\eqref{eq:objective} and the preservation of speech fidelity (see Section~\ref{sec:attack_model}).

The filter used for evaluation is selected according to a measure of the 
attack success rate for the full set of utterances in Part~1.  The 
attack success rate is defined as the fraction of spoofed utterances for which
$\mathbf{N}(\cm{\spf * \advflt}) > 0.5$, where $\mathbf{N}(\cdot)$ normalises CM scores to reflect probabilities in $[0,1]$.
This setting reflects the point at which the CM considers $\spf$ more likely to be bona fide than spoofed.\footnote{In practice, $\mathbf{N(\cdot)}$ is implemented as a Softmax operation applied to the activations of the final linear layer of each CM model.}
Our specific implementation is available as open-source and can be used to reproduce our results under the same GPU environment.\footnote{\url{github.com/eurecom-asp/malafide}}

\subsection{Countermeasures}
\label{sec:cms}
We used three different CM systems to verify the effectiveness of the adversarial filter attack. They are described below.  All are available as open-source.

\newpara{\textbf{RawNet2}}~\cite{jung2020improved}, an end-to-end (E2E) model developed originally for ASV, has also been applied to spoofing and deepfake detection~\cite{tak2021rawnet}.\footnote{\url{github.com/eurecom-asp/rawnet2-antispoofing}}
The first network layer 
is a bank of 20 mel-scaled sinc filters, and is convolved directly with raw waveform inputs. 
The sinc-layer is followed by a series of six residual blocks and a gated recurrent unit (GRU) which produces a score indicative of whether the input is bona fide or spoofed. 

\newpara{\textbf{AASIST}}\footnote{\label{fn:assist}\url{github.com/clovaai/aasist}}\cite{jung2022aasist} is a state-of-the-art E2E spoofing CM solution based upon the RawNet2 CM described above.
It uses the same sinc-layer and residual network to extract higher-level feature representations. 
The back-end includes a spectro-temporal graph attention network (RawGAT-ST)~\cite{tak2021end}, heterogeneous graph attention layers and max graph operations to integrate temporal and spectral representations. 
Scores are generated using a readout operation and a fully connected output layer.

\newpara{\textbf{Self-supervised leaning (SSL)}} based front-ends have gained increasing attention in a range of speech-related tasks in recent years, including for spoofing and deepfake detection~\cite{wang2021investigating,eom2022anti,wang2022investigating,martin2022vicomtech}. 
The SSL-based CM architecture \cite{tak2022automatic}\footnote{\url{github.com/TakHemlata/SSL_Anti-spoofing}} is a two-stage model with SSL-based feature extraction and a back-end comprising graph attention and pooling layers, a single fully-connected layer and an output layer.  
It is the only of the three CMs used in this work which is trained using external data beyond that permitted by ASVspoof evaluation rules.  Nonetheless, SSL solutions have improved substantially on the previous state of the art~\cite{tak2022automatic,wang22_ssl}.
The SSL feature extractor is a pre-trained wav2vec 2.0 model~\cite{babu2021xls}\footnote{\url{github.com/pytorch/fairseq/tree/main/examples/wav2vec/xlsr}} the weights of which are fine-tuned during CM training.

\subsection{Metrics}
All results are reported as equal error rate (EER) estimates and are obtained using the standard SASV evaluation protocol~\cite{jungsasv2022}.  Results reported in Section~5 are CM EERs computed using a mix of bona fide and spoofed trials.  
Results reported in Secion~6 are SASV-EERs computed 
using a mix of target (positive class) and both non-target and spoofed utterances (negative class).

\section{Results}
        
    \begin{table*}[!t]
        \caption{
        %Evaluation SPF-EERs of filters applied on different CM with unseen data. The first row indicates the filter name, and the second row indicates the tested CM system.
        CM performance in terms of the EER (\%) in both white-box and black-box settings.  Results shown without filtering and with Malafide filtering using filters of different lengths $L$ (number of taps).
        %.s of filters applied on different CM with unseen data. The first row indicates the filter name, and the second row indicates the tested CM system.
        }
        % \vspace{2mm}
        \centerline{
        \resizebox{\textwidth}{!}{%
        \renewcommand{\arraystretch}{1}
        \setlength\tabcolsep{2pt}
        \begin{tabular}{ccccccccccc}
        %\hline
        %\texttt{Column 1} & \texttt{2} & \texttt{3} &  \texttt{4 } &  \texttt{ 5} &  \texttt{ 6} &  \texttt{7 } &  \texttt{ 8} &  \texttt{ 9} &  \texttt{ 10} \\
       % \hline
        \hline
            & \multicolumn{3}{c}{\textbf{AASIST Malafide Attack}} & \multicolumn{3}{c}{\textbf{RawNet2 Malafide Attack}}& \multicolumn{3}{c}{\textbf{SSL Malafide Attack}}  \\
        \cmidrule(lr){2-4} \cmidrule(lr){5-7} \cmidrule(lr){8-10} 
        \multirow{1}{*}{}  &   \multicolumn{1}{c}{AASIST} & \multicolumn{1}{c}{RawNet2} & \multicolumn{1}{c}{SSL}& \multicolumn{1}{c}{AASIST} & \multicolumn{1}{c}{RawNet2} & \multicolumn{1}{c}{SSL}& \multicolumn{1}{c}{AASIST} & \multicolumn{1}{c}{RawNet2} & \multicolumn{1}{c}{SSL}\\
        % \cmidrule(lr){1-1}
        \multirow{1}{*}{\textbf{Filter length}}&   \multicolumn{1}{c}{(white-box)} & \multicolumn{1}{c}{(black-box)} & \multicolumn{1}{c}{(black-box)}& \multicolumn{1}{c}{(black-box)} & \multicolumn{1}{c}{(white-box)} & \multicolumn{1}{c}{(black-box)}& \multicolumn{1}{c}{(black-box)} & \multicolumn{1}{c}{(black-box)} & \multicolumn{1}{c}{(white-box)}\\
        \hline

        \multirow{1}{*}{no filter}   & \multicolumn{1}{c}{0.71} & \multicolumn{1}{c}{3.29} & \multicolumn{1}{c}{1.01} & \multicolumn{1}{c}{0.71} & \multicolumn{1}{c}{3.29} & \multicolumn{1}{c}{1.01} & \multicolumn{1}{c}{0.71} & \multicolumn{1}{c}{3.29} & \multicolumn{1}{c}{1.01}\\
        \hline
        \multirow{1}{*}{65}   & \multicolumn{1}{c}{5.54} & \multicolumn{1}{c}{8.94} & \multicolumn{1}{c}{3.63} & \multicolumn{1}{c}{2.35} & \multicolumn{1}{c}{15.59} & \multicolumn{1}{c}{1.76} & \multicolumn{1}{c}{0.07} & \multicolumn{1}{c}{10.73} & \multicolumn{1}{c}{8.33}\\
        \hline
        \multirow{1}{*}{129}   & \multicolumn{1}{c}{8.15} & \multicolumn{1}{c}{10.21} & \multicolumn{1}{c}{1.76} & \multicolumn{1}{c}{1.57} & \multicolumn{1}{c}{20.71} & \multicolumn{1}{c}{1.83} & \multicolumn{1}{c}{0.04} & \multicolumn{1}{c}{12.00} & \multicolumn{1}{c}{9.53}\\
        \hline
        \multirow{1}{*}{257}   & \multicolumn{1}{c}{9.73} & \multicolumn{1}{c}{9.72} & \multicolumn{1}{c}{0.97} & \multicolumn{1}{c}{1.05} & \multicolumn{1}{c}{21.46} & \multicolumn{1}{c}{1.16} & \multicolumn{1}{c}{0.22} & \multicolumn{1}{c}{9.53} & \multicolumn{1}{c}{6.81}\\
        \hline
        \multirow{1}{*}{513}   & \multicolumn{1}{c}{13.87} & \multicolumn{1}{c}{11.18} & \multicolumn{1}{c}{0.19} & \multicolumn{1}{c}{0.93} & \multicolumn{1}{c}{21.95} & \multicolumn{1}{c}{0.97} & \multicolumn{1}{c}{0.08} & \multicolumn{1}{c}{11.25} & \multicolumn{1}{c}{6.98}\\
        \hline
        \multirow{1}{*}{1025}   & \multicolumn{1}{c}{12.71} & \multicolumn{1}{c}{15.81} & \multicolumn{1}{c}{0.15} & \multicolumn{1}{c}{1.05} & \multicolumn{1}{c}{21.91} & \multicolumn{1}{c}{0.19} & \multicolumn{1}{c}{0.04} & \multicolumn{1}{c}{10.54} & \multicolumn{1}{c}{12.30}\\
        \hline
        \multirow{1}{*}{2049}   & \multicolumn{1}{c}{9.36} & \multicolumn{1}{c}{23.93} & \multicolumn{1}{c}{0.26} & \multicolumn{1}{c}{1.68} & \multicolumn{1}{c}{16.19} & \multicolumn{1}{c}{0.12} & \multicolumn{1}{c}{0.15} & \multicolumn{1}{c}{8.98} & \multicolumn{1}{c}{10.91}\\
        \hline
        \multirow{1}{*}{4097}   & \multicolumn{1}{c}{6.62} & \multicolumn{1}{c}{19.18} & \multicolumn{1}{c}{1.80} & \multicolumn{1}{c}{1.27} & \multicolumn{1}{c}{11.92} & \multicolumn{1}{c}{0.12} & \multicolumn{1}{c}{0.23} & \multicolumn{1}{c}{8.42} & \multicolumn{1}{c}{11.15}\\
        \hline

        \end{tabular}}
        }
        \label{tab:results}
        \end{table*}

Results are presented in Table~\ref{tab:results}. 
They show EERs without and with the use of different length adversarial filters (column 1).
Results are also shown for filters optimised for one CM (row 1) and tested against another (row 2); column 2 shows EERs for a filter optimised for, and tested with the AASIST CM (a white-box setting), whereas column 3 shows EERs for a filter optimised for the AASIST CM but tested with the RawNet2 CM (a black-box setting).  

Results for the three white-box settings shown in columns~2, 6 and 10 of Table~\ref{tab:results} show that adversarial filtering provokes substantial increases in the CM EER, with the greatest EERs being achieved with filters of either 513 coefficients (AASIST and RawNet2) or 1025 coefficiences (SSL).
With a maximum EER of 12.3\%, the SSL CM is the most robust.  
The most vulnerable is the RawNet2 CM for which the maximum EER is 22.0\%, although it also has the highest initial EER without adversarial filtering of~3.3\%.

We now turn to black-box settings.
Columns 3 and 4 of Table~\ref{tab:results} show black-box results for filters learned using the AASIST CM.  
The attack is transferable to 
the RawNet2 CM (maximum EER of 23.9\%), but less so to the SSL CM (maximum EER of 3.6\%). 
Black-box results for filters learned using the RawNet2 CM are shown in columns 5 and 7 of 
Table~\ref{tab:results}. 
The attack is still effective for both AASIST and SSL CMs, with maximum EERs of approximately 2\% for relatively shorter length filters.
Black-box results for the SSL CM shown in columns 8 and 9 of Table~\ref{tab:results} show that the attack transfers to the RawNet2 CM (EER as high as 12\%) but not to the AASIST CM, for which the attack is wholly unsuccessful. 
All CMs are vulnerable to adversarial filtering attacks under white-box settings and, albeit to a lesser extent, also black-box settings. 
The RawNet2 CM is particularly vulnerable across all conditions.

\section{Impact upon ASV}

    \begin{table}[!t]
        \caption{SASV-EER (\%) results for the score-level fusion  %of using the proposed filter to attack both ASV and CM systems. Results are SASV-EERs based on the score fusion of 
        of ASV and CM sub-systems under Malafide attacks.
        }
        % \vspace{2mm}
        \centerline{
        \resizebox{\columnwidth}{!}{%
        \renewcommand{\arraystretch}{1.1}
        \begin{tabular}{lcccc}
        \hline
        % \multirow{2}{*}{\textbf{System}} & \multicolumn{3}{c}{\textbf{SASV-EER}}  \\
        % \cmidrule(lr){2-4}
        \multirow{2}{*}{CM} & \multirow{2}{*}{no filter} & \multicolumn{3}{c}{Malafide Attack}\\
        \cmidrule(lr){3-5}
         &  & AASIST & RawNet2 & SSL \\
        \hline
        AASIST & 1.23 & 11.21 & 0.82 & 0.82 \\
        % AASIST w/o filter & 1.32 & 1.32 & 1.32 \\
        % AASIST w/ filter  & 11.55 & 0.82 & 0.82 \\
        \hline
        RawNet2 & 2.62 & 6.91 & 6.96 & 3.99 \\
        % RawNet2 w/o filter & 2.72 & 2.72 & 2.72  \\
        % RawNet2 w/ filter & 6.91 & 7.47 & 3.99 \\
        \hline
        SSL & 1.46 & 1.46 & 1.46 & 1.57 \\
        % SSL w/o filter & 1.66 & 1.66 & 1.66  \\
        % SSL w/ filter & 1.46 & 1.46 & 1.77  \\
        
        \hline
        \end{tabular}}
        }
        \label{tab:sasv}
        \end{table} 

Results presented thus far show that the proposed adversarial filter attack can be used to compromise the reliability of a CM subsystem.
Here we show that the attack preserves speech fidelity so that it is also successful in compromising 
both CM and ASV subsystems -- the ASV subsystem by the spoofing attack, and the CM subsystem by the Malafide filter attack.

Results presented in Table~\ref{tab:sasv} show spoofing-aware speaker verification (SASV) EERs~\cite{jungsasv2022} obtained using fused CM and ASV scores computed from the same set of trials used for independent CM evaluation (Section 5).
SASV-EERs are shown without filtering (column 2) and with Malafide filters (columns 3-5) learned using one of the three CMs and tested under the same mix of white-box and black-box settings as Table~1 (column 1).
All Malafide filter results are for a 257-tap filter which provokes the highest averaged SASV-EERs.
Under all white-box settings, the SASV-EER increases, to 11.2\% for the AASIST CM, 7.0\% for the RawNet2 CM and to 1.6\% for the SSL CM.  
The trend for black-box settings is similar to that for independent CM results, with the system that uses the SSL CM being wholly robust, but with the system that uses the RawNet2 CM being universally vulnerable.

Informal listening tests revealed that shorter-length filters better preserve speech fidelity whereas longer-length filters result in detectable reverberation. 
Longer-length filters degrade speech fidelity to the point that spoofing attacks are no longer successful in compromising the ASV system.

We acknowledge that SASV-EERs are heavily dependent on the proportion of negative class trials that are spoofed (as opposed to non-target).
In this respect, performance estimates are not necessarily indicative of what might be expected in the wild where, for instance, spoof attacks may be less prevalent.

\section{Conclusions}
The work reported in this paper shows that the reliability of spoofing countermeasures (CMs) can be compromised using adversarial, linear time-invariant filters and that these can also be configured to compromise integrated CM and automatic speaker verification (ASV) systems. 
Malafide attacks are a threat in both white-box and black-box settings, and a RawNet2 CM is particularly vulnerable.

Results for integrated CM and ASV systems
show that Malafide attacks are successful in manipulating a spoofing CM and, when used in conjunction with spoofing attacks and by introducing only modest perturbations, ASV subsystems too.  
The performance of the integrated system that uses a self-supervised learning (SSL) CM is an exception; performance is reasonably robust.  This is likely caused by the introduction of greater distortion to the speech signal that is needed to compromise the relatively more complex CM.  The same distortion interferes with the ability of the spoofing attack to compromise the ASV system.
Future work should study similar convolutional attacks that are optimised to compromise both CM \emph{and} ASV subsystems.  These attacks might expose vulnerabilities of even SSL-based systems.  Such work and is critical if we are to protect confidence in the reliability of voice biometrics technology.

\section{Acknowledgements}
The second author is supported by the TReSPAsS-ETN project funded by the European Union’s Horizon 2020 research and innovation programme under the Marie Skłodowska-Curie grant agreement No.\ 860813. The third author is supported by the VoicePersonae project funded by the French Agence Nationale de la Recherche (ANR) and the Japan Science and Technology Agency (JST).

\balance
\bibliographystyle{IEEEtran}
\bibliography{mybib}

\begin{thebibliography}{10}
\providecommand{\url}[1]{#1}
\csname url@samestyle\endcsname
\providecommand{\newblock}{\relax}
\providecommand{\bibinfo}[2]{#2}
\providecommand{\BIBentrySTDinterwordspacing}{\spaceskip=0pt\relax}
\providecommand{\BIBentryALTinterwordstretchfactor}{4}
\providecommand{\BIBentryALTinterwordspacing}{\spaceskip=\fontdimen2\font plus
\BIBentryALTinterwordstretchfactor\fontdimen3\font minus
  \fontdimen4\font\relax}
\providecommand{\BIBforeignlanguage}[2]{{%
\expandafter\ifx\csname l@#1\endcsname\relax
\typeout{** WARNING: IEEEtran.bst: No hyphenation pattern has been}%
\typeout{** loaded for the language `#1'. Using the pattern for}%
\typeout{** the default language instead.}%
\else
\language=\csname l@#1\endcsname
\fi
#2}}
\providecommand{\BIBdecl}{\relax}
\BIBdecl

\bibitem{tolosana2019biometricPAD}
R.~Tolosana, M.~Gomez-Barrero, C.~Busch, and J.~Ortega-Garcia, ``{Biometric
  presentation attack detection: Beyond the visible spectrum},'' \emph{IEEE
  Transactions on Information Forensics and Security}, vol.~15, pp. 1261--1275,
  2019.

\bibitem{george2019FacePAD}
A.~George, Z.~Mostaani, D.~Geissenbuhler, O.~Nikisins, A.~Anjos, and S.~Marcel,
  ``{Biometric face presentation attack detection with multi-channel
  convolutional neural network},'' \emph{IEEE Transactions on Information
  Forensics and Security}, vol.~15, pp. 42--55, 2019.

\bibitem{yamagishi2021_ASV_spoof}
J.~Yamagishi, X.~Wang, M.~Todisco, M.~Sahidullah, J.~Patino, A.~Nautsch,
  X.~Liu, K.~A. Lee, T.~Kinnunen, N.~Evans, and H.~Delgado, ``{ASV}spoof 2021:
  Accelerating progress in spoofed and deepfake speech detection,'' in
  \emph{Proc. {ASV}spoof 2021 Workshop}, 2021, pp. 47--54.

\bibitem{jungsasv2022}
J.-w. Jung, H.~Tak, H.-j. Shim, H.-S. Heo, B.-J. Lee, S.-W. Chung, H.-J. Yu,
  N.~Evans, and T.~Kinnunen, ``{SASV 2022: The first spoofing-aware speaker
  verification challenge},'' in \emph{Proc. Interspeech 2022}, 2022, pp.
  2893--2897.

\bibitem{wang22_ssl}
X.~Wang and J.~Yamagishi, ``{Investigating self-supervised front ends for
  speech spoofing countermeasures},'' in \emph{Proc. Speaker Odyssey Workshop},
  2022, pp. 100--106.

\bibitem{tak2022automatic}
H.~Tak, M.~Todisco, X.~Wang, J.-w. Jung, J.~Yamagishi, and N.~Evans,
  ``Automatic speaker verification spoofing and deepfake detection using
  wav2vec 2.0 and data augmentation,'' in \emph{Proc. Speaker Odyssey
  Workshop}, 2022, pp. 112--119.

\bibitem{jung2022aasist}
J.-w. Jung, H.-S. Heo, H.~Tak, H.-j. Shim, J.~S. Chung, B.-J. Lee, H.-J. Yu,
  and N.~Evans, ``{AASIST: Audio anti-spoofing using integrated
  spectro-temporal graph attention networks},'' in \emph{Proc. ICASSP 2022},
  2022, pp. 6367--6371.

\bibitem{cm_adv_atks1}
S.~Liu, H.~Wu, H.-y. Lee, and H.~Meng, ``Adversarial attacks on spoofing
  countermeasures of automatic speaker verification,'' in \emph{2019 IEEE
  Automatic Speech Recognition and Understanding Workshop (ASRU)}, 2019, pp.
  312--319.

\bibitem{black_box_cm_additive}
Y.~Zhang, Z.~Jiang, J.~Villalba, and N.~Dehak, ``{Black-box attacks on spoofing
  countermeasures using transferability of adversarial examples},'' in
  \emph{Proc. Interspeech 2020}, 2020, pp. 4238--4242.

\bibitem{gomezalanis2021adversarial}
A.~Gomez-Alanis, J.~A. Gonzalez, and A.~M. Peinado, ``{Adversarial
  transformation of spoofing attacks for voice biometrics},'' in \emph{Proc.
  IberSPEECH 2021}, 2021, pp. 255--259.

\bibitem{universal_perturbations_CM}
X.~Zhang, X.~Zhang, W.~Liu, X.~Zou, M.~Sun, and J.~Zhao, ``{Waveform level
  adversarial example generation for joint attacks against both automatic
  speaker verification and spoofing countermeasures},'' \emph{Engineering
  Applications of Artificial Intelligence}, vol. 116, p. 105469, 2022.

\bibitem{adv_kassis}
A.~Kassis and U.~Hengartner, ``{Practical attacks on voice spoofing
  countermeasures},'' \emph{arXiv preprint arXiv:2107.14642}, 2021.

\bibitem{wu2020defense}
H.~Wu, S.~Liu, H.~Meng, and H.-y. Lee, ``{Defense against adversarial attacks
  on spoofing countermeasures of ASV},'' in \emph{Proc. ICASSP 2020}, 2020, pp.
  6564--6568.

\bibitem{neekhara19b_universal}
P.~Neekhara, S.~Hussain, P.~Pandey, S.~Dubnov, J.~McAuley, and F.~Koushanfar,
  ``{Universal adversarial perturbations for speech recognition systems},'' in
  \emph{Proc. Interspeech 2019}, 2019, pp. 481--485.

\bibitem{xie2020real}
Y.~Xie, C.~Shi, Z.~Li, J.~Liu, Y.~Chen, and B.~Yuan, ``{Real-time, universal,
  and robust adversarial attacks against speaker recognition systems},'' in
  \emph{Proc. ICASSP 2020}, 2020, pp. 1738--1742.

\bibitem{adv_attacks_goodfellow}
I.~J. Goodfellow, J.~Shlens, and C.~Szegedy, ``Explaining and harnessing
  adversarial examples,'' in \emph{3rd International Conference on Learning
  Representations, {ICLR} 2015, San Diego, CA, USA, May 7-9, 2015, Conference
  Track Proceedings}, Y.~Bengio and Y.~LeCun, Eds., 2015.

\bibitem{audio_adv_attacks_carlini}
N.~Carlini and D.~Wagner, ``Audio adversarial examples: Targeted attacks on
  speech-to-text,'' in \emph{2018 IEEE Security and Privacy Workshops (SPW)},
  2018, pp. 1--7.

\bibitem{universal_perturbations_image}
S.-M. Moosavi-Dezfooli, A.~Fawzi, O.~Fawzi, and P.~Frossard, ``Universal
  adversarial perturbations,'' in \emph{Proc. of the IEEE Conference on
  Computer Vision and Pattern Recognition (CVPR)}, July 2017.

\bibitem{universal_perturbations_ASR}
P.~Neekhara, S.~Hussain, P.~Pandey, S.~Dubnov, J.~McAuley, and F.~Koushanfar,
  ``{Universal adversarial perturbations for speech recognition systems},'' in
  \emph{Proc. Interspeech 2019}, 2019, pp. 481--485.

\bibitem{unviersal_perturbations_ASV}
W.~Zhang, S.~Zhao, L.~Liu, J.~Li, X.~Cheng, T.~F. Zheng, and X.~Hu, ``{Attack
  on practical speaker verification system using universal adversarial
  perturbations},'' in \emph{Proc. ICASSP 2021}, 2021, pp. 2575--2579.

\bibitem{universal_audio_attacks_ASC}
S.~Abdoli, L.~G. Hafemann, J.~Rony, I.~B. Ayed, P.~Cardinal, and A.~L. Koerich,
  ``{Universal adversarial audio perturbations},'' \emph{arXiv preprint
  arXiv:1908.03173}, 2019.

\bibitem{he_init}
K.~He, X.~Zhang, S.~Ren, and J.~Sun, ``Delving deep into rectifiers: Surpassing
  human-level performance on imagenet classification,'' in \emph{2015 IEEE
  International Conference on Computer Vision (ICCV)}, 2015, pp. 1026--1034.

\bibitem{wang2020ASVspoof}
X.~Wang, J.~Yamagishi, M.~Todisco, H.~Delgado, A.~Nautsch, N.~Evans,
  M.~Sahidullah, V.~Vestman, T.~Kinnunen, K.~A. Lee \emph{et~al.}, ``{ASV}spoof
  2019: A large-scale public database of synthetized, converted and replayed
  speech,'' \emph{Computer Speech \& Language}, vol.~64, 2020, 101114.

\bibitem{kingma2014adam}
D.~P. Kingma and J.~L. Ba, ``{Adam: A method for stochastic optimization},'' in
  \emph{Proc. of the 3rd International Conference on Learning Representations
  (ICLR)}, 2015.

\bibitem{jung2020improved}
J.-w. Jung, S.-b. Kim, H.-j. Shim, J.-h. Kim, and H.-J. Yu, ``{Improved
  {R}awNet with filter-wise rescaling for text-independent speaker verification
  using raw waveforms},'' in \emph{Proc. Interspeech 2020}, 2020, pp.
  1496--1500.

\bibitem{tak2021rawnet}
H.~Tak, J.~Patino, M.~Todisco, A.~Nautsch, N.~Evans, and A.~Larcher,
  ``{End-to-end anti-spoofing with RawNet2},'' in \emph{Proc. ICASSP
  2021}.\hskip 1em plus 0.5em minus 0.4em\relax IEEE, 2021, pp. 6369--6373.

\bibitem{tak2021end}
H.~Tak, J.~Jung, and J.~Patino, ``{End-to-end spectro-temporal graph attention
  networks for speaker verification anti-spoofing and speech deepfake
  detection},'' in \emph{Proc. {ASV}spoof 2021 workshop}, 2021, pp. 1--8.

\bibitem{wang2021investigating}
X.~Wang and J.~Yamagishi, ``{Investigating self-supervised front ends for
  speech spoofing countermeasures},'' in \emph{Proc. Speaker Odyssey Workshop},
  2022, pp. 100--106.

\bibitem{eom2022anti}
Y.~Eom, Y.~Lee, J.~S. Um, and H.~Kim, ``{Anti-spoofing using transfer learning
  with variational information bottleneck},'' in \emph{Proc. Interspeech 2022},
  2022, pp. 3568--3572.

\bibitem{wang2022investigating}
X.~Wang and J.~Yamagishi, ``Investigating active-learning-based training data
  selection for speech spoofing countermeasure,'' in \emph{the 2022 IEEE Spoken
  Language Technology Workshop}, 2022, pp. 585--592.

\bibitem{martin2022vicomtech}
J.~M. Mart{\'\i}n-Do{\~n}as and A.~{\'A}lvarez, ``The vicomtech audio deepfake
  detection system based on wav2vec2 for the 2022 add challenge,'' in
  \emph{Proc. ICASSP 2022}, 2022, pp. 9241--9245.

\bibitem{babu2021xls}
A.~Babu, C.~Wang, A.~Tjandra, K.~Lakhotia, Q.~Xu, N.~Goya, K.~Singh, P.~{von
  Platen}, Y.~Saraf, J.~Pino \emph{et~al.}, ``{XLS-R}: Self-supervised
  cross-lingual speech representation learning at scale,'' in \emph{Proc.
  Interspeech 2022}, 2022, pp. 2278--2282.

\end{thebibliography}

\end{document}